# A near-field scanned microwave probe for spatially localized electrical metrology

Vladimir V. Talanov, [a] André Scherz, Robert L. Moreland, and Andrew R. Schwartz

*Neocera, Inc., 10000 Virginia Manor Road, Beltsville, MD 20705*

**Abstract:** We have developed a near-field scanned microwave probe with *a sampling volume* of approximately 10 µm in diameter, which is the smallest one achieved in near-field microwave microscopy. This volume is defined to confine close to 100% of the probe's net sampling reactive energy, thus making the response virtually independent on the sample properties outside of this region. The probe is formed by a 4 GHz balanced stripline resonator with a few-micron tip size. It provides non-contact, non-invasive measurement and is uniquely suited for spatially localized electrical metrology applications, e.g. on semiconductor production wafers.

---

[a] Electronic mail: talanov@neocera.com

The electrodynamic response of near-field scanned microwave probes (NSMPs) is fundamentally due to reactive energy, electric and/or magnetic, stored in the near-field penetrating the sample [1]. The volume confining *half* of this energy has been used to theoretically estimate the spatial (or *imaging*) resolution of modern near-field microwave microscopes [2, 3]. For apertureless ones with a sharp STM- or AFM-like tip [2-4] this *imaging volume* was found to be on the order of the tip apex curvature, depending on the sample permittivity and tip-sample distance. Experimentally, the spatial resolution down to few nanometers has been demonstrated [4], when defined as the smallest feature that can be *imaged* on a high contrast sample. Since the *imaging volume* contains only fraction (e.g. 50%) of the net sampling energy, the NSMP response depends on and can be even dominated by the sample properties *outside* of this region [5, 6]. It has been observed that the near-field *outside* the imaging volume, often referred to as the stray field (or capacitance), affects the response [5, 6]. For example, such "non-tip-end" *E*-field accounts for up to half of the frequency shift $\Delta F$ in a widely used NSMP formed by the sharpened wire terminating a coaxial resonator [6]. Since, $\Delta F/F \approx \Delta W/W$ for small perturbations in a resonant cavity [7] ($W$ is the total energy stored in the resonator), about half of the sampling reactive energy is stored in this parasitic field. Various structures have been proposed in the literature to reduce or shield those fields [5].

Generally speaking, to *quantitatively* interpret the measurement on any inhomogeneous (laterally or depth-wise) sample one must to consider the *entire* volume the probe response is collected from. To address this issue, we introduce the definition of a NSMP *sampling volume* confining close to 100% of the net sampling reactive energy. While significant attention has been paid to improving the NSMPs imaging resolution (see [1] and references therein) by reducing, in effect, their imaging volume, there is much to be learned about controlling the sampling volume. This is crucial for applications such as electrical metrology on semiconductor production wafers or combinatorial materials libraries, which require the measurement to be virtually insensitive to the sample properties outside a certain area, e.g. a test-key in the 80-µm-wide wafer scribe-line or a single cell in the library. Since the probe is much smaller than the radiation wavelength, the near-field has a static character. Therefore, it extends over a length-scale governed by the "characteristic" probe dimension, such as the length of the entire sharpened wire or AFM cantilever typically exceeding a few hundred microns. Hence, the two volumes, the imaging and sampling, can differ by as much as a few orders of magnitude in linear size for the same probe, thus making apertureless NSMPs inapplicable for *quantitative* measurements (viz. *imaging*) on many practical test structures or small samples. While their imaging resolution on high permittivity sample can be even much smaller than the tip apex because of the field enhancement due to the induced image of the tip [3], the sampling volume is more than a few hundred microns in size as governed by the probe characteristic dimension. Here we present a NSMP with the sampling volume/area of about 10 µm in diameter. Rather than attempting to shield the stray fields inherently present in unbalanced *dipole*-type geometries such as a sharpened wire or AFM cantilever, we employ a different approach where a balanced transmission stripline forms a *quadrupole*-like probe.

Our probe is fabricated from a quartz bar of 1×1 mm$^2$ in cross-section pulled down to sub-micron size using a $CO_2$ laser micropipette puller. A tapered parallel strip transmission line with no cut-off frequency is formed by depositing ~3-µm-thick Al onto two opposite sidewalls of the pulled bar. Throughout the entire taper the bar maintains a square cross-section, yielding a line with nearly uniform characteristic impedance $Z_0$~100Ω. To form a well-defined electrically open tip, the tapered end is trimmed up to 7–10 µm in size using a micropipette beveler (Fig.1). To increase the measurement sensitivity, a λ/2 parallel strip resonator (PSR) is formed by etching the Al strips to a length of ~2.5 cm. A magnetic loop couples the microwave radiation into the resonator. The PSR and the coupling loop are installed inside a metallic enclosure with the taper protruding a few millimeters out via a clear hole in the enclosure wall (Fig.2). Besides the parallel strip (i.e. *balanced odd*) eigen-mode with a resonant frequency *F*~4 GHz and unloaded quality-factor *Q*~100, this structure also supports a coaxial-like (*unbalanced even*) mode at ~6 GHz with *Q*~1000 similar to the one employed in coaxial NSMPs with a sharp tip [2, 3, 5, 6].

We operate our probe in the *balanced odd mode* only, where the protruding portion of PSR forms an electrically small quadrupole-like antenna with the currents being close to zero at the tip and linearly dependent on *z*. Its near-zone field (or reactive energy) is mostly confined in between the Al strips, while the parasitic far-field radiated power is a few orders of magnitude less than that of unbalanced dipole-like geometries [2, 3, 5, 6]. Full 3D high-frequency finite element modeling [8] shows that at the tip the sampling *E*-field forms a well-confined "cloud" with a characteristic dimension on the order of the tip size *D* (see Fig.2, insets). This field is similar to the fringe field of a *D*-thick parallel plate capacitor or a *D*-long electric dipole parallel to *x*. It will be experimentally confirmed below that this dimension defines the sampling volume for our probe. One can see that despite more than two orders of magnitude difference in $\varepsilon$ the depth of field penetration as well as its lateral extend are nearly the same. When a tip is brought into close proximity to the sample the cloud "penetrates" it, the energy stored in the sampling *E*-field reduces by an amount dependent on the sample permittivity, and the probe resonant frequency *F* decreases. By applying the impedance transformation to a fore-opened λ/2 resonator [7], the relative change in the probe resonant frequency for a low loss sample is:

$$\frac{\Delta F}{F} \approx \frac{Z_0 (X_{t2} - X_{t1})}{\pi X_{t1} X_{t2}} \qquad (1)$$

where $X_t$ is the tip reactance, and $X_t >> Z_0$. $X_t$ can be found using the Pointing theorem for energy conservation [7] or a lumped element model [1], e.g. $X_t = -1/\omega C_t$, where $C_t$ is the tip capacitance and $\omega = 2\pi F$. From Eq. (1) the probe sensitivity to $C_t$ is $\sim 10^{-19}$ Farad for typical 0.1 ppm precision in *F* measurement. It is performed with the aid of a voltage controlled oscillator (VCO), which is locked onto the resonance (i.e. a minimum of the PSR |$S_{11}$|) using a frequency-tracking loop similar to [3]. The common parasitic impact of the VCO's thermal drift is avoided in our setup by using a high-stability frequency counter (Fig. 2) to measure the loop carrier equal to *F*.

The tip-sample distance control for our probe is based on a shear-force (SF) method [9]. The quartz bar forms a mechanical resonator with a fundamental frequency ~3 kHz and *Q*-factor ~ few hundred, which is excited by dithering the enclosure at this frequency with nanometer amplitude using a piezo tube (Fig. 2). The tip is illuminated with a laser beam projecting onto a photo-detector that ac output depends on the tip vibration amplitude, which is, in turn, a strong function of the tip-sample distance. This signal is fed via a lock-in amplifier and PID controller into a piezo z-stage holding the probe, to control the tip-sample separation with precision ~2 nm. Assuming the tip is flat and parallel to the sample surface, the effective tip-sample operating distance is estimated to be between 50 and 100 nm based on the typical tip geometry (see Fig. 1) and observed resonant frequency shift ~ 1 MHz. Given the inevitable imperfections in the tip form there is likely a point that is significantly closer to the sample (e.g. ~10 nm), which actually provides for the SF interaction. This is the first implementation of an *optically detected* SF in near-field microwave microscopy. The developed distance control is precise, non-contact and virtually independent of the sample electrodynamic properties. The laser beam does not interfere with the microwaves, unlike the alternate tuning-fork oscillator approach [10] where the tuning-fork is mechanically attached to the probe tip. The piezo z-stage is mounted onto a mechanical z-stage supported by a gantry bridge. Wafers up to 300 mm in diameter are scanned with a 350 by 350 mm travel xy-stage beneath the probe. The apparatus sits on a vibration-isolated platform inside an environmental chamber at ambient temperature.

Now we experimentally estimate the volume confining >99% of the probe sampling electrical energy in order to determine its sampling volume in the "worst" case of a low permittivity material under test with a high permittivity background. To find the vertical extent of the sampling volume an air ($\varepsilon$=1) film with a metallic backing (low-resistivity <5 mΩ·cm Si wafer, $|\varepsilon|$>>1) is employed. Fig. 3a shows the probe measured resonant frequency shift $\Delta F(z)=F_e-F(z)$ vs. the normalized distance $z/D$ from the wafer surface to the tip (i.e. the air film thickness). $F_e$ is the probe frequency with no sample present, i.e. $F_e=F(z\rightarrow\infty)$. To find it we assume that for $z/D$>>1 $\Delta F(z)$ should fall off faster than $1/z$ to keep the net sampling energy finite since $\Delta F \propto \Delta W$. Thus, $F_e$ was estimated (with precision ~10 kHz) by empirically fitting the data to a linear combination of $1/z^{1.1}$ and a five-term exponential decay, and extrapolating $z$ to infinity. For $z/D$>1 $\Delta F$ changes by less than 1% of the total frequency shift $F_e-F(z=0)$. Therefore, according to perturbation theory [7], less than 1% of the net sampling energy is stored below the ~1×$D$-thick layer beneath the tip.

The lateral extent of the sampling volume was measured via a line-scan across the edge of a 60×60 μm² Cu patch buried under a 414-nm-thick so-called low-$k$ dielectric film ($k=\varepsilon_{film}$ is the film dielectric constant). Fig. 3b shows the result when the Al strips are parallel to the patch edge (the scan with Al strips perpendicular to the edge scan looks similar). The lateral position $x$ is normalized by the tip size $D$. Using an approach similar to [11] the measured frequency shift was converted into $\varepsilon_{film}$. Again, outside of a transition range ~1.5×$D$ long the "as measured" $\varepsilon_{film}$ exhibits no change. Above the patch we obtained $\varepsilon_{film}$=3.21±0.01, which is close to the nominal value of 3.15. Note, that outside the patch the "as measured" $\varepsilon_{film}$ exhibits non-physical value ≤1 because in the data analysis the film thickness was fixed at 414 nm, which is incorrect outside the patch (see Fig. 3b). The above results confirm that due to the static nature of the near-field both the vertical and lateral extents of the sampling volume are on the order of the tip characteristic dimension $D$, e.g. the sampling volume for our probe roughly forms a cylinder of height ~$D$ and diameter ~1.5×$D$ located just beneath the tip and oriented along $z$.

To conclude, we have reported on a near-field scanned microwave probe achieving the smallest *sampling volume* (or area) in near-field microwave microscopy ~10 µm. It can be reduced even further by decreasing the tip size or by increasing the multipole order of the antenna forming the probe (e.g., a magnetic quadrupole proposed by Osofsky and Schwarz [12]). The probe-sample interaction could potentially be affected by exciting the surface waves in the sample, which was not observed in our experiments. The advantages of our NSMP are: a) non-contact, non-invasive measurement; b) the probe is fabricated entirely from quartz and aluminum, which makes it non-contaminating; and c) no electrical contact to or grounding of the sample under test is required since both probing electrodes are located above the sample and capacitively coupled to it. The probe applications to spatially localized electrical metrology, such as quantitative characterization of low-*k* interconnect dielectrics on semiconductor *production* wafers, will be published elsewhere.

This work was partially supported by NSF-SBIR 0078486 and NIST-ATP 70NANB2H3005. We thank Dr. H. Christen, Prof. I. Smolyaninov and Dr. B. Ming for technical assistance, and Prof. T. Venky Venkatesan and Prof. S. Anlage for valuable discussions.

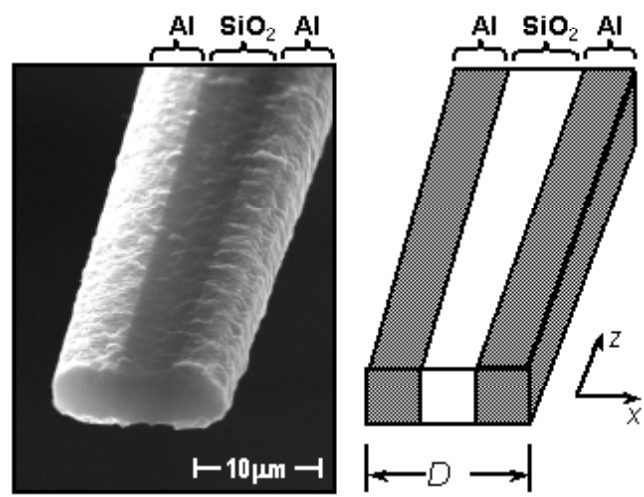

**Fig.1**. SEM image and 3D sketch of a typical parallel strip resonator tip formed of two Al strips deposited onto a tapered quartz bar.

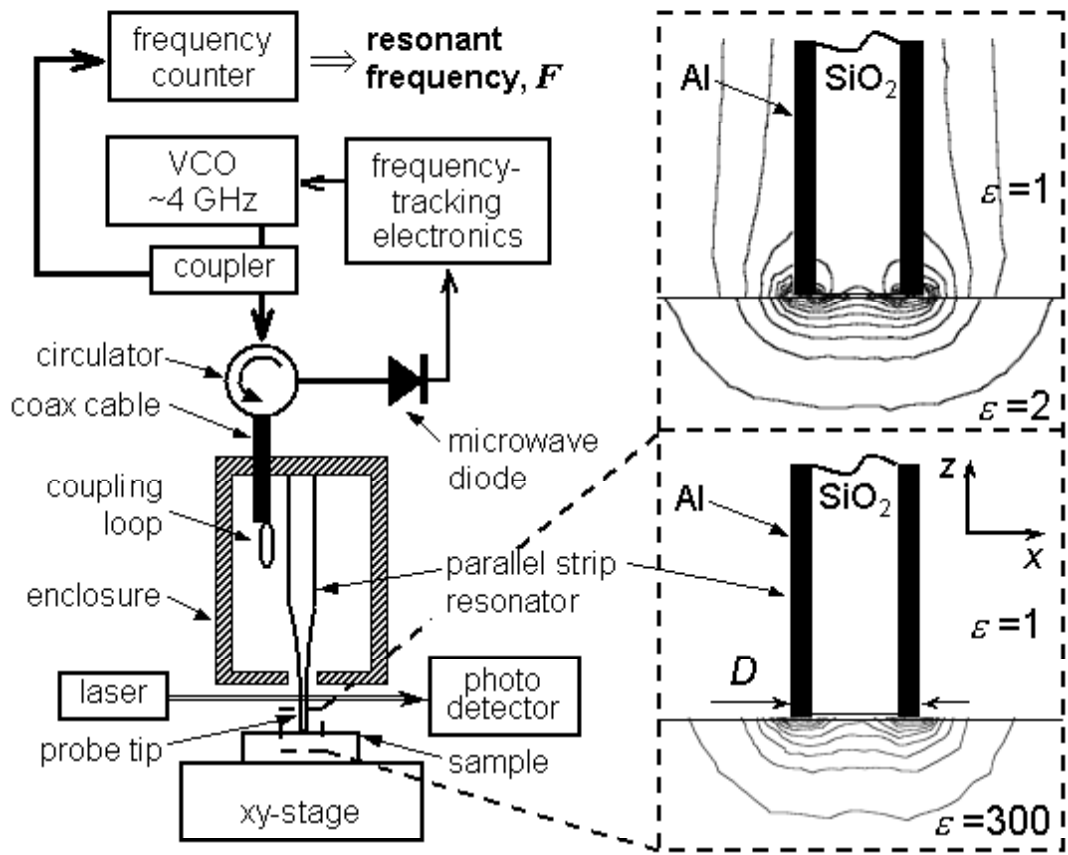

**Fig.2.** Apparatus schematic showing the probe, electronics, and shear-force setup. Note, that in reality the laser beam is *perpendicular* to the direction of enclosure dithering. **Insets**: intensity contour plots of the fringe $E$-field in the $xz$-plane for bulk samples with $\varepsilon=2$ (top) and $\varepsilon=300$ (bottom); in the latter the contours above the sample are not shown for clarity; on both plots the outermost contour line shows where the sampling field magnitude drops 20 times from its value just under the sample surface; the tip-sample distance is 100 nm, $D=4$ $\mu$m.

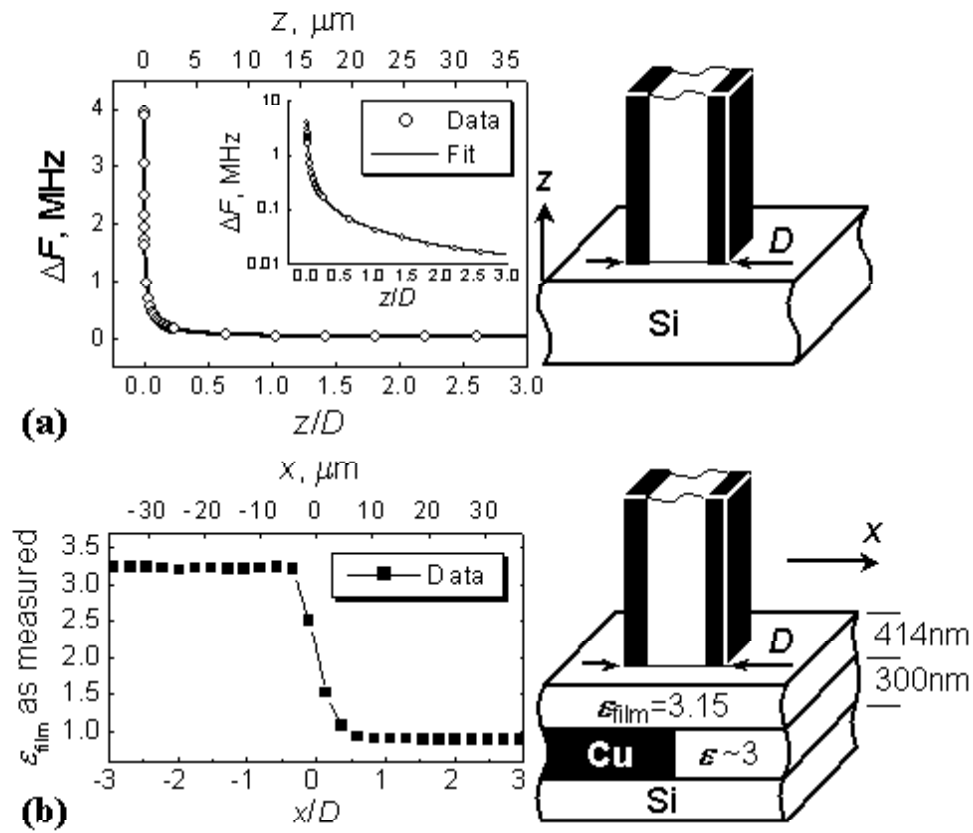

**Fig.3.** (a) Measured frequency shift $\Delta F$ vs. normalized tip-sample distance $z/D$ for a low-resistivity Si wafer on linear and (**inset**) semi-log scales; solid line is the empirical fit (see text); $z=0$ is the shear-force distance; axis on top is absolute $z$. Tip size $D\sim 12$ $\mu$m. (b) Film dielectric constant as measured vs. normalized position $x/D$ across the edge of a Cu patch located at $x<0$ and buried under a 414-nm-thick dielectric film with nominal $\varepsilon=3.15$; solid line is a guide to the eye; axis on top is absolute $x$. Tip size $D\sim 12$ $\mu$m.